\documentclass[twocolumn,showpacs,preprintnumbers,amsmath]{revtex4}

\usepackage[usenames]{color}
\usepackage{graphicx}
\usepackage{epsfig}
\usepackage{dcolumn}
\usepackage{bm}
\usepackage{latexsym}
\usepackage{amsfonts}
\usepackage{amssymb}

\newcommand{\beq}{\begin{equation}}
\newcommand{\eeq}{\end{equation}}
\newcommand{\beqa}{\begin{eqnarray}}
\newcommand{\eeqa}{\end{eqnarray}}
\newcommand{\bea}{\begin{array}}
\newcommand{\ena}{\end{array}}

\begin{document}
\title{Radionic Non-uniform Black Strings}
\author{Takashi Tamaki}
\email{tamaki@tap.scphys.kyoto-u.ac.jp}
\author{Sugumi Kanno}
\email{sugumi@tap.scphys.kyoto-u.ac.jp}
\author{Jiro Soda}
\email{jiro@tap.scphys.kyoto-u.ac.jp}
\affiliation{Department of Physics, Kyoto University,
606-8501, Japan}

\date{\today}
\begin{abstract}
Non-uniform black strings in the two-brane system are investigated 
using the effective action  approach.  It is shown that the radion 
acts as a non-trivial hair of the black strings. From the brane point of view, 
the black string appears as the deformed dilatonic black hole which becomes 
dilatonic black hole in the single brane limit and reduces to the 
Reissner-Nordstr\"om black hole  in the close limit of two-branes. 
The stability of solutions is demonstrated using the catastrophe theory. 
From the bulk point of view, the black strings are proved to be non-uniform. 
Nevertheless, the zeroth law of black hole thermodynamics still holds. 
\end{abstract}

\preprint{KUNS-1861}

\maketitle
\section{Introduction}
Recent progress in superstring theory has suggested the possibility
 that we resides on the hypersurface in the higher dimensional
 space-time~\cite{Horava}. This so-called  braneworld scenario has rapidly
 become popular since Randall and Sundrum proposed simple models 
 where we live in a brane embedded in AdS$_5$ space-time~\cite{Randall}. 
 There are two types of Randall-Sundrum (RS) scenarios. 
 In the first RS model (RS1), the two flat branes are embedded 
in AdS$_5$ space-time to solve the hierarchy problem. In the second RS 
model (RS2), one positive tension brane is embedded in  AdS$_5$ space-time. 

 The cosmology and  black holes in RS models have been investigated 
 actively~\cite{others:Cos,Hawking,Shinkai,Shibata,Emparan,Kudoh,others:BH}. 
 The common issue is to understand 
 effects of the bulk geometry on the braneworld.
 The  radion and Kaluza-Klein modes are key ingredients for this aim. Here,
 it should be noted that most of important phenomena occur at low energies or 
 the low curvature regime.  At low energies, the radion would dominate 
 the effects of the extra  dimension.  In the case of the RS2 model, 
however, the radion does not exist. Hence, except for the small contribution 
from the Kaluza-Klein modes, the conventional Einstein theory is recovered at 
 low energies in the RS2 model. On the other hand, 
  in the case of the RS1  model, the existence of the radion is crucial.
 Recently, we have developed a systematic method to study the 
 braneworld at low energies using the gradient expansion method~\cite{Kanno}.
 In particular, we have shown that the gravity in the RS1 model can be described 
 by a 4 dimensional  scalar-tensor theory  and the information of the bulk 
 can be reconstructed through holograms. 
 The cosmology has been investigated using this approach and found an 
 interesting scenario~\cite{born}.  As our method is quite general, 
 it is  also applicable to black hole physics. 

In the conventional general relativity, Schwarzschild solution is the 
simplest stable black hole solution. 
While, in the superstring theory, dilatonic black holes are 
found by Gibbons and   Maeda, and independently by Garfinkle, Horowitz, 
and Strominger (GM-GHS)~\cite{GM-GHS}. 
Interestingly,  dilatonic black holes are stable 
in the same sense as the Schwarzschild black hole~\cite{Wilczek}. 
In this sense, these are realistic black holes in the superstring theory. 
Let us go back to the braneworld. 
The conventional Schwarzschild black hole  can be realized as a 
section of the uniform black string.  However, this is not a unique 
possibility. The black hole can be localized on the brane, 
although it may not appear as an exact Schwarzschild black hole. Still, to 
find all of the stable black holes in the braneworld is an open issue. 
From the string theoretic point of view, it is also  desirable to 
find GM-GHS like solutions in the braneworld scenario. 
Unfortunately, no exact black string solution is known in this case. 
Therefore, it is interesting to seek black strings corresponding to 
GM-GHS black holes in RS models.

In the RS2 model, it is argued that the black hole 
should be localized on the brane because of ``Gregory-Laflamme" 
instability of the black string~\cite{Gregory}. 
This localized black holes have been investigated
intensively~\cite{Hawking,Shinkai,Shibata,Emparan,Kudoh}. 
On the other hand, in the RS1 model, the black  string exists without 
suffering from the  Gregory-Laflamme instability if the distance 
between two branes is less than the radius of the black hole on the brane. 
In the opposite condition, Gregory-Laflamme instability commences. 
Intriguingly, since the dynamics of the radion controls the 
length of the black string, it can trigger  the transition 
from the stable  to unstable black string. It is argued that 
the fate of the unstable black string is either the localized black hole or 
the non-uniform black string~\cite{Kengo,Wiseman,Choptuik}. 
Clearly, the first step to this direction is to understand the role of 
the radion in the black strings.  
Although numerical analysis is practical to attack this issue, 
the effort must be made toward the analytic understanding of the phenomena.
 
The purpose of this paper is  to present a semi-analytic approach 
to investigate  black strings in the RS1 model. 
We investigate the black strings corresponding to GM-GHS black holes 
on the brane using the proposed method. In particular, the role of the 
radion in the non-uniform black strings is explored. 
   
 This paper is organized as follows. In Sec.~II, we briefly review our  
 effective action approach. In Sec.~III, we describe  basic equations
 for the numerical analysis.  In Sec.~IV,  
we describe our solution from the 4-dimensional point of view
 with an emphasis on the role of the radion. In Sec.~V, 
 we  view our solution from the bulk. In Sec.~VI, 
 we summarize our results.

\section{Effective Action Approach}
We begin by reviewing the low energy effective action 
derived in previous papers~\cite{Kanno,kanno1}. 
We consider an $S_1/Z_2$ orbifold space-time with the two branes 
as the fixed points. In the RS1 model, 
the positive ($\oplus$) and the negative ($\ominus$) tension branes 
are embedded in AdS$_5$ with the curvature radius $\ell$. 
Our system is described by the action 
\begin{eqnarray}
S&=&\frac{1}{2\tilde{\kappa}^2}\int d^5 x \sqrt{-g}\left({\cal R}
+\frac{12}{\ell^2}\right)
\nonumber\\
&&-\sum_{i=\oplus,\ominus}\sigma_i 
\int d^4 x \sqrt{-g^{i\mathrm{\hbox{-}brane}}}  \nonumber  \\
&&+\sum_{i=\oplus,\ominus} 
\int d^4 x \sqrt{-g^{i\mathrm{\hbox{-}brane}}}
\,{\cal L}_{\rm matter}^i \ ,\label{5D:action}
\end{eqnarray}
where ${\cal R}$, $g^{i\mathrm{\hbox{-}brane}}_{\mu\nu}$, $\sigma_{i}$ 
and $\tilde\kappa^2$ are the 5-dimensional scalar curvature, the induced 
metric on the $i$-brane, the tension on the $i$-brane, 
and the 5-dimensional gravitational constant, respectively. 
Here, we assume the relations $\sigma_\oplus=6/(\tilde{\kappa}^2\ell)$ and 
$\sigma_\ominus=-6/(\tilde{\kappa}^2\ell)$. 

For general nonflat branes, we cannot keep both the two branes straight in 
the Gaussian normal coordinate system. Hence, we use the following 
coordinate system to describe the geometry of the brane model: 
\begin{equation}
ds^{2}=e^{2\eta (x^{\mu})}dy^{2}+g_{\mu\nu}(y,x^{\mu})dx^{\mu}dx^{\nu}\ . 	
\label{coordinate}
\end{equation}
We place the branes at $y=0$ ($\oplus$-brane) and $y=\ell$ ($\ominus$-brane) 
in this coordinate system. In this coordinate system, the proper distance
 between two branes with fixed $x^\mu$ is represented as
\begin{eqnarray}
 d(x^\mu ) = \int^\ell_0 e^{\eta (x^\mu )} dy \ .
\end{eqnarray}
Hence, we call $\eta$ the radion. 

The strategy we take in this paper is the following.
Let us start with the variation of the action
\begin{eqnarray}
  \delta S [g_{\mu\nu}, h^i_{\mu\nu}]
  = \frac{\delta S}{\delta g_{\mu\nu} } \delta g_{\mu\nu} 
     + \sum_{i=\oplus,\ominus}
     \frac{\delta S}{\delta h^i_{\mu\nu} } \delta h^i_{\mu\nu} =0 \ .
\end{eqnarray}
The variation with respect to $g_{\mu\nu}$ gives the bulk Einstein equations
 and the variation with respect to $h_{\mu\nu}$ yields the junction conditions.
First, we solve the bulk equations of motion with $h^i_{\mu\nu}$ fixed,
 then we get the relation
\begin{eqnarray}
  g_{\mu\nu} = g_{\mu\nu}[h^i_{\mu\nu}] \ .
  \label{bulk-metric}
\end{eqnarray}
Substituting Eq.(\ref{bulk-metric}) into junction conditions, we get the 
4-dimensional 
effective equations of motion for the induced metric $h^i_{\mu\nu}$. 
Or, by substituting Eq.(\ref{bulk-metric}) into the action, 
we obtain 4-dimensional effective action   
\begin{eqnarray}
 S_{\rm eff} = S[g_{\mu\nu} [h^i_{\mu\nu}] , h^i_{\mu\nu}] \ .
 \label{eff-action}
\end{eqnarray}
From Eq.(\ref{eff-action}), the 4-dimensional effective equations of motion 
can be reproduced. 

Of course, we must resort to some approximation method 
to solve bulk equations of motion. 
 As the usual phenomena occurs at low energies  
\begin{equation}
\frac{\rho_i}{|\sigma_i|} 
	\sim \frac{\tilde{\kappa}^2\rho_i}{\tilde{\kappa}^2 |\sigma_i|}\,
	\sim (\frac{\ell}{L})^2 \ll 1 \ ,
	\label{rel:low-energy}
\end{equation}
where $L$ denotes the characteristic length scale of the curvature  
on the brane, we can employ the gradient expansion method~\cite{tomita}. 
We seek the metric as a perturbative series 
with the number of derivatives increasing with the order 
of iteration, that is, $ O((\ell/L)^{2n})$, $n=0,1,2,\cdots$ as
\begin{eqnarray}
\hspace{-5mm}    	  g_{\mu\nu} (y,x^\mu ) &=&
  		b^2 (y,x) \left[ h_{\mu\nu} (x^\mu) 
  		+ g^{(1)}_{\mu\nu} (y,x^\mu)
      		+ \cdots  \right] , \label{expansion:metric} \\
      		b(y,x)&=&\exp (-ye^{\eta}/\ell), 
\end{eqnarray}
where we put the Dirichlet boundary condition 
$g_{\mu\nu} (y=0, x) =  h_{\mu\nu} (x)$ at the $\oplus$-brane. 
 Substituting (\ref{expansion:metric}) into the bulk Einstein equations, we can
 obtain the relation (\ref{bulk-metric}) perturbatively. 
 The first order correction is given by
\begin{eqnarray}
\hspace{-9mm}&&g_{\mu\nu}^{(1)}=-\frac{\ell^{2}}{2}\left[
\left(\frac{1}{b^{2}}-1\right)
\left(R_{\mu\nu}-\frac{1}{6}h_{\mu\nu}R\right)\right.
\nonumber  \\
\hspace{-9mm}&&-\left(\frac{1}{b^{2}}-1-
\frac{2ye^{\eta}}{\ell}\frac{1}{b^{2}}\right)
\left(\eta_{|\mu\nu}+\frac{h_{\mu\nu}}{2}\eta^{|\alpha}\eta_{|\alpha}\right)
\nonumber  \\
\hspace{-9mm}&&
\left.+\frac{2y^{2}e^{2\eta}}{\ell^{2}b^{2}}
\left(\eta_{|\mu}\eta_{|\nu}-
\frac{h_{\mu\nu}}{2}\eta^{|\alpha}\eta_{|\alpha}\right)
+\left(\frac{1}{b^{4}}-1\right)\chi_{\mu\nu}\right],
\label{correction}	
\end{eqnarray}
where
\begin{eqnarray}
\chi^{\mu}{}_{\nu}&=&-\frac{\kappa^2(1-\Psi)}{2 \Psi} 
	\left\{ T^{\oplus\mu}{}_{\nu} 
	+ (1-\Psi)T^{\ominus\mu}{}_{\nu}\right\}  \nonumber\\
&&	-\frac{1}{2 \Psi} \left[ \left(  \Psi^{|\mu}{}_{|\nu} 
	-\delta^\mu_\nu  \Psi^{|\alpha}{}_{|\alpha} \right) 
	\right. \nonumber \\
&&	\left.+\frac{3}{2(1 -\Psi )} \left( \Psi^{|\mu}  \Psi_{|\nu}
  	-\frac{1}{2} \delta^\mu_\nu  \Psi^{|\alpha} \Psi_{|\alpha} 
  	\right) \right],     
  	\label{A:chi}
\end{eqnarray}
is nothing but projected Weyl tensor $E_{\mu\nu}$~\cite{ShiMaSa}.
Here, $\kappa^{2}:=\tilde{\kappa}^{2}/\ell$ and 
\begin{eqnarray}
\Psi := 1- \exp (-2 e^\eta)\ ,
\label{dandrad} 
\end{eqnarray}
is defined for convenience. We also call $\Psi$ the radion.
 The junction condition gives the effective equations of motion
 from which the action for the $\oplus$-brane can be read off as
\begin{eqnarray}
\hspace{-5mm}S_{\rm\oplus}&=&\frac{1}{2 \kappa^2} \int d^4 x \sqrt{-h} 
	\left[ \Psi R (h) - \frac{3}{2(1- \Psi )} 
     	\Psi^{|\alpha} \Psi_{|\alpha} \right] \nonumber\\
    &&	+ \int d^4 x \sqrt{-h} {\cal L}^\oplus 
      	+ \int d^4 x \sqrt{-h} \left(1-\Psi \right)^2 {\cal L}^\ominus .
      	\label{A:action} 
\end{eqnarray}
If we can obtain a solution for this system (\ref{A:action}), the bulk metric 
corresponding to the 4-dimensional effective theory is  given by
\begin{equation}
  	g_{\mu\nu} = (1-\Psi )^{y/\ell} \left[
       		h_{\mu\nu} (x) + g^{(1)}_{\mu\nu} (h_{\mu\nu} , \Psi, 
        	T^\oplus_{\mu \nu } ,  T^\ominus_{\mu \nu } , y)  \right] \ , 
        	\label{holograms}
\end{equation}
which corresponds to Eq.(\ref{bulk-metric}). 
Thus the bulk metric is completely determined by the 
energy momentum tensors on both branes, the radion and the induced 
metric on the $\oplus$-brane. Therefore, once the 4-dimensional solution of 
the quasi-scalar-tensor gravity is given, one can reconstruct the bulk 
geometry from these data.  The quasi-scalar-tensor gravity (\ref{A:action}) 
works as  holograms at the low energy.

\section{Basic Equations for Numerical Analysis}
Below, we consider a dilaton field coupled to the electromagnetic field 
on the $\oplus$-brane since it appears in the compactification process. 
Then, we have 
\begin{eqnarray}
\hspace{-3mm}S_{\rm\oplus}&=&\frac{1}{2 \kappa^2} \int d^4 x \sqrt{-h} 
	\left[ \Psi R (h) - \frac{3}{2(1- \Psi )} 
     	\Psi^{|\alpha} \Psi_{|\alpha} \right] \nonumber\\
&&-\int d^4 x \sqrt{-h} \left[\frac{1}{2}\phi^{|\alpha}\phi_{|\alpha}
+\frac{1}{4}e^{-2a\phi}F_{\mu\nu}F^{\mu\nu}\right],  
\label{A:action2} 
\end{eqnarray}
where $\phi$ and $a$ are a dilaton field and its coupling 
constant, respectively. 
Since our theory is a scalar-tensor type theory, we call
this original action the Jordan-frame effective action.
In order to discuss the property of radion, it is sometimes convenient to 
move to the Einstein frame in which the action takes the canonical
Einstein-scalar form. 
Applying a conformal transformation 
$h_{\mu\nu}=\frac{1}{\Psi}q_{\mu\nu}$ and 
introducing a new field 
\begin{eqnarray}
\psi&=&-\sqrt{\frac{3}{2\kappa^2}}\log\left|\frac{\sqrt{1-\Psi}-1}{\sqrt{1-\Psi}
+1}\right| \ ,  
\label{A:field}
\end{eqnarray}
we obtain the Einstein-frame effective action 
\begin{eqnarray}
\hspace{-5mm}S_{\rm\oplus}&=&\int d^4 x \sqrt{-q}
	\left[\frac{1}{2\kappa^2}R(q)-\frac{1}{2}
	(\nabla\psi)^2\right.
	\nonumber\\
&&\left.-\frac{1}{2}\left(\cosh^2{\frac{\kappa}{\sqrt{6}}\psi}\right)
	(\nabla\phi)^2-\frac{1}{4}e^{-2a\phi}F^2 
	\right],\label{eqn:action}
\end{eqnarray}
where $\nabla$ denotes the covariant derivative with respect to 
the metric $q_{\mu\nu}$. We also call $\psi$ the radion. 
We take this action and consider the static and 
spherically symmetric metric as 
\beqa
ds^{2}=-f(r) e^{-2\delta(r)}dt^{2}+f(r)^{-1}dr^{2}
+r^{2}d\Omega^{2},
\label{eqn:metric}
\eeqa
where $f(r):=1-\kappa^{2}m(r)/4\pi r$. We construct a black hole on the brane 
and construct a  black string using holograms (\ref{holograms}). 
We choose a vector potential $A_{\mu}$ as 
\beqa
A_{\mu}=(0,0,0,Q_{m}\cos\theta),
\label{eqn:gauge}
\eeqa
where $Q_{m}$ is a magnetic charge. If we also consider an electric charge, our 
system has an electric-magnetic duality $F \to e^{-2a\phi}\tilde{F}$, 
$\phi \to -\phi$, where $\tilde{F}$ is the Hodge dual of $F$. 
Hence we can easily obtain the results for the 
electrically charged case. Our basic equations are 
\beqa
\hspace{-5mm}m'&=&4\pi\left(\frac{r^{2}}{2}fU+\frac{Q_{m}^{2}e^{-2a\phi}}{2r^{2}}\right), 
\label{eqn:einstein}   \\
\hspace{-5mm}\delta'&=&
 -\frac{\kappa^{2}}{2}rU,  \\     
\hspace{-5mm}\phi''&=&-\frac{2\kappa}{\sqrt{6}}\psi '\phi '\tanh(\frac{\kappa}{\sqrt{6}}\psi)
\nonumber  \\
\hspace{-5mm}&&-\frac{1}{f}\left[\phi 'V
+\frac{ae^{-2a\phi}Q_{m}^{2}}{r^{4}\cosh^{2}(\frac{\kappa}{\sqrt{6}}\psi)}\right] ,
\label{eqn:dilaton}  \\
\hspace{-5mm}\psi''&=&-\frac{\psi'V}{f}+
\frac{\kappa}{\sqrt{6}}(\phi')^{2}\cosh(\frac{\kappa}{\sqrt{6}}\psi)
\sinh(\frac{\kappa}{\sqrt{6}}\psi)\ ,
\label{eqn:radion}
\eeqa
where the prime denotes the derivative with respect to $r$ and
\beqa
U&:=&(\psi')^{2}+(\cosh^{2}\frac{\kappa}{\sqrt{6}}\psi)(\phi')^{2}\ ,  
\label{s}   \\
V&:=&\frac{2}{r}-\frac{\kappa^{2}m}{4\pi r^{2}}-\frac{\kappa^{2} Q_{m}^{2}}{2r^{3}}e^{-2a\phi} 
\label{t}  \ .
\eeqa

We assume the existence of a regular 
event horizon at $r=r_{H}$. So we have 
\beqa
m_{H}&=&\frac{4\pi r_{H}}{\kappa^{2}},\;\; \delta_H,\ \phi_{H},\ 
\psi_H< \infty ,\ \psi_{H}'=0\ ,
\label{mth}  \\
\phi_{H}'&=&-\frac{ae^{-2a\phi_{H}}Q_{m}^{2}}{
V_{H}r_{H}^{4}\cosh^{2}(\frac{\kappa}{\sqrt{6}}\psi_{H})}\ .
\label{ath}
\eeqa
The variables with subscript $H$ are evaluated 
at the horizon.
The boundary conditions at spatial infinity 
to guarantee asymptotic flatness are 
\begin{eqnarray}
m(\infty) =: M=const.,\ \ \delta (\infty)=0, \nonumber \\ 
\phi(\infty)=0,\ \ \psi(\infty)=:\psi_{\infty}=const.\ , 
\label{atinf}
\end{eqnarray}
where $M$ corresponds to a gravitational mass of the black hole. 
We will obtain the black hole 
solutions numerically by solving Eqs.(\ref{eqn:einstein}) - (\ref{eqn:radion})
iteratively with boundary conditions 
(\ref{mth}), (\ref{ath}) and (\ref{atinf}). 

\section{View from the Brane}

The radion is affected by the matter on the brane and varies depending on 
the place on the brane. 
 The other way around, the configuration of matter  is affected by the radion. 
 Hence, the radion must play an important role in the system of black string. 
 As to the GM-GHS black holes, their stability is already known.
 It would be interesting to investigate if the radion destroys the stability
 or not.  

\subsection{Radion as a Hair}
If we include the moduli fields, the first law of black hole thermodynamics 
has the term depending on the scalar charge~\cite{GKK}. 
 The contribution from both the dilaton field and the radion field can be 
 expected. We shall examine it starting from the action (\ref{eqn:action}). 
 Here, we consider the 
spherically symmetric and the magnetically charged case for simplicity. 
We denote the time translational Killing vector as 
$k:=\frac{\partial}{\partial t}$ and the null vector  orthogonal to 
the event horizon $H$ as $n_{\mu}$ which is normalized as 
\beqa
n_{\mu}k^{\mu}=-1\ .
\label{normal}
\eeqa
We represent the volume element and the surface element as $d\sigma_{\mu}$ 
and $dS_{\mu\nu}$, respectively. Then, the gravitational mass $M$ and the 
``magnetic charge" $Q_{m}$ can be defined as 
\beqa
M&:=&\frac{-1}{4\pi}\int_{\infty}k^{\mu ;\nu}dS_{\mu\nu}\ ,\ \\
Q_{m}&:=&-\int_{\infty}e^{-2a\phi}F^{\mu\nu}dS_{\mu\nu}\ .
\label{integral}
\eeqa
We can obtain the mass formula using the standard procedure~\cite{Wald} as
\beqa
M&=&\frac{1}{\kappa^{2}}\int \left[R-(\nabla\psi)^{2}
-\cosh^{2}(\frac{\kappa}{\sqrt{6}}\psi)(\nabla\phi)^{2}\right. \nonumber \\
&&\left.-\frac{1}{2}e^{-2a\phi}F^{2}\right]k^{\mu}d\sigma_{\mu} 
+\frac{T_{H}}{2}A+2\Phi_{H}Q_{H}\ ,
\label{mass-formula}
\eeqa
where $T_{H}$, $\Phi_{H}$ and $Q_{H}$ are the Hawking temperature, the 
magnetic potential at the horizon and the ``magnetic charge" at the horizon, 
respectively. Taking the variation of this equation, 
we obtain the first law of black hole thermodynamics, 
\beqa
\delta M&=& T_{H}\frac{\delta A}{4}+\Phi_{H}\delta Q_{H}  
-\Sigma\delta \psi \ ,
\label{dmass-formula}
\eeqa
where we defined the radionic charge $\Sigma$ as 
\beqa
\Sigma:=2\int_{\infty}\psi^{;\mu}k^{\alpha}dS_{\alpha\mu}\ .
\eeqa

\begin{figure}[htbp]
\psfig{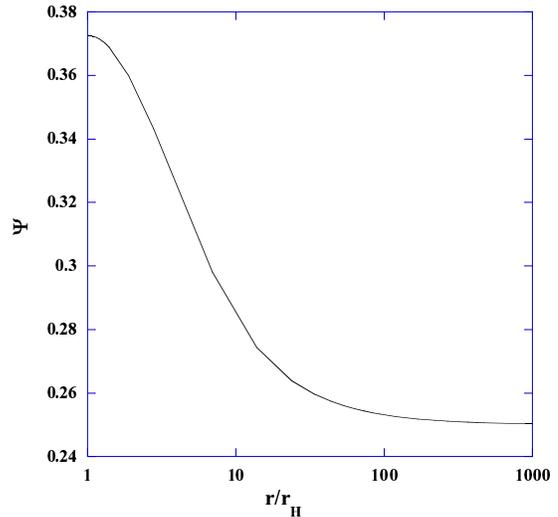}
\caption{The behavior of $\Psi$ for $\Psi_{\infty}=0.25$, $\lambda_{H}=0.119$ 
and $\bar{a}=\sqrt{3}$. \label{r-Psi} }
\end{figure}

This formula shows that the dilatonic charge can not be a true hair, 
because it is absorbed into the definition of 
the ``magnetic charge"~\cite{Okai}. 
However, the radion field is a hair of black strings. 
To derive (\ref{dmass-formula}), we used the property that $\psi$ 
asymptotically behaves as $O(1/r)$ which can be  confirmed 
by asymptotic analysis of Eqs.~(\ref{eqn:einstein})-(\ref{eqn:radion}). 
We can prove that $\psi$ ($\Psi$) is monotonically 
increasing (decreasing) function of $r$ from Eq.~(\ref{eqn:radion}). 
In Fig.~\ref{r-Psi}, as an example, $\Psi$  for a solution 
$\bar{a}:=\sqrt{2}a/\kappa=\sqrt{3}$, $\Psi_{\infty}:=\Psi (\infty)=0.25$ 
and the horizon radius $\lambda_{H}:=\sqrt{2}r_{H}/\kappa Q_{m}=0.119$ 
is depicted~\cite{foot}. 

\subsection{Stability of the Black String}
First, we summarize the relation between $\Psi$ (or $\psi$) and $d$ using 
(\ref{dandrad}) and (\ref{A:field}) in Table~\ref{table}. $\Psi$ ($\psi$) is 
a monotonically increasing (decreasing) function with respect to $d$. 

\begin{table}[htbp]
\begin{center}
\caption{Relation between variables  \label{table}   }
\tabcolsep=5mm
\begin{tabular}{l|l|l|l}
 & close limit & & single brane limit \\
\hline
$d$    &$0$      &$\to$&$\infty$   \\
\hline            
$\psi$&$\infty$&$\to$&$0$  \\
\hline
$\Psi$&$0$      &$\to$&$1$ \\
\end{tabular}
\end{center}
\end{table}

In the single brane limit $\psi\to 0$, the effect of the radion ceases.
 Hence, the solution approaches GM-GHS solution. 
In the close limit $\psi\to\infty$, we have $\phi (r)=0$ from 
Eqs.~(\ref{eqn:dilaton}) and (\ref{ath}). 
Thus, only Reissner-Nordstr\"om (RN) solutions are possible in this limit 
 because of the no-hair theorem~\cite{Bekenstein}. 
Therefore, we expect that the non-trivial radion 
($\psi\neq 0,\infty$) interpolates  RN and  GM-GHS solutions. 

\begin{figure}[htbp]
\psfig{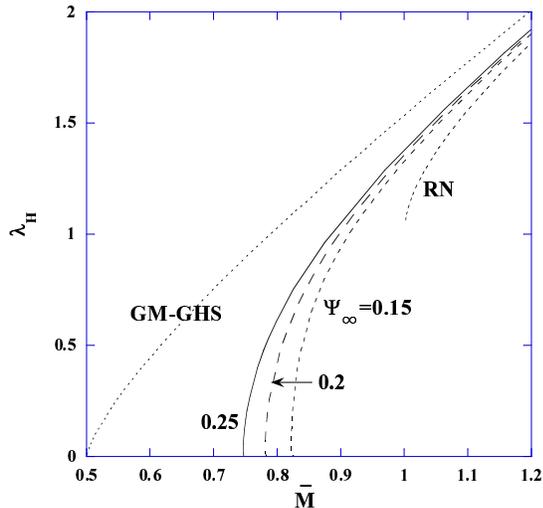}
\caption{The relation between the horizon radius $\lambda_{H}$ and 
the gravitational mass $\bar{M}$ for $\bar{a}=\sqrt{3}$. 
\label{Figart3} }
\end{figure}

As an illustration, we show the relation between the horizon radius 
$\lambda_{H}$ and the gravitational 
mass $\bar{M}:=\frac{\kappa M}{4\pi \sqrt{2}Q_{m}}$ 
in Fig.~\ref{Figart3}. 
 In fact, in the single brane limit, the solution approaches GM-GHS and 
the effect of the dilaton field fades away in the close limit. 

We also notice that solutions exist until $\lambda_{H}\to 0$ except 
for the RN solution. This suggests that there is no inner horizon. 
Indeed, we can prove it. First, rewrite 
Eq. (\ref{eqn:dilaton}) as 
\beqa
[fr^{2}e^{-\delta}\cosh^{2}(\frac{\kappa}{\sqrt{6}}\psi)\phi']'
=\frac{a}{r^{2}}e^{-2a\phi}e^{-\delta}Q_{m}^{2}\ .
\eeqa
It follows that 
$D:=fr^{2}e^{-\delta}\cosh^{2}(\frac{\kappa}{\sqrt{6}}\psi)\phi'$ 
monotonically increases with $r$ for $a>0$. Hence, $D<0$ holds
inside the horizon and 
then $f$ cannot become $0$, which implies that there is no inner horizon. 
Thus, the causal structure  is the same as that of Schwarzschild black hole. 
This suggests that our solutions are stable as the 
GM-GHS solutions~\cite{Wilczek}. 

\begin{figure}[htbp]
\psfig{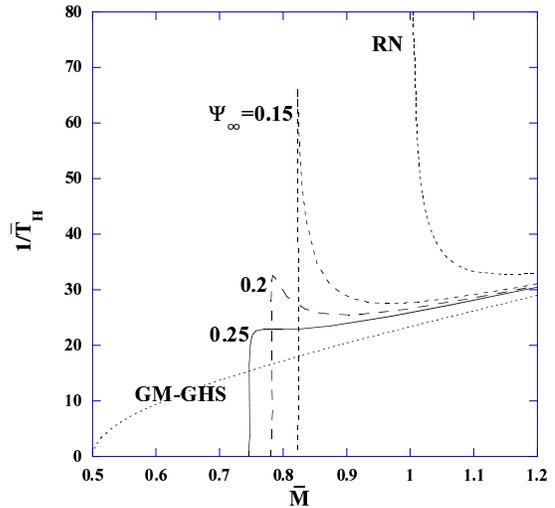}
\caption{$\bar{M}$-$1/\bar{T}_{H}$ for $\bar{a}=\sqrt{3}$. 
\label{Figart3t} }
\end{figure}

We can also argue stability of our solutions using the catastrophe theory. 
In Fig.~\ref{Figart3t}, the relation between the inverse temperature 
$1/\bar{T}_{H}:=(1/T_{H})/(\kappa Q_{m}/2)$ and $\bar{M}$ 
 are plotted.  According to the catastrophe theory, the stability changes at 
$d(1/T_{H})/dM=\infty$~\cite{Katz,catas,Tamaki} ( see Appendix.~A). 
 Since we cannot find the point $d(1/T_{H})/dM=\infty$ in the graph
 for various parameters,  our solutions are stable in the catastrophic sense. 

\begin{figure}[htbp]
\psfig{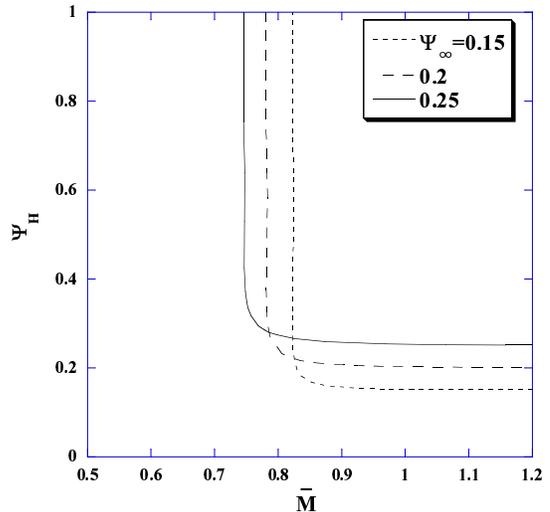}
\caption{$\bar{M}$-$\Psi_{H}$ for $\bar{a}=\sqrt{3}$. 
\label{M-Psih}}
\end{figure}

Finally, let us interpret a feature seen in Fig.~\ref{Figart3t}. 
Seeing the relation between $\bar{M}$ and $\Psi_{H}$ for fixed $\Psi_\infty$ 
in Fig.~\ref{M-Psih}, we find that $\Psi_{H}$ varies from the small value 
to $1$ by reducing the mass. Thus, the separation between branes at 
the horizon is small for large mass, hence, 
the solution mimics the RN black hole. By reducing the mass, the separation 
goes to the infinity.  Thus, $T_{H}$ eventually diverges like 
the GM-GHS solution as seen in Fig.~\ref{Figart3t}. 


\section{View from the bulk}

Our interest here is the $y$-dependence of the horizon which had not been 
investigated so far in the RS1 model. The hologram (\ref{holograms}) 
 can be used to see the shape of the black strings.  
Differences from the previous papers~\cite{Kengo,Wiseman,Choptuik} are 
that the matter is confined on the $\oplus$-brane and the branes are deformed 
by black strings. These differences will affect the shape of the horizon. 
Indeed, we will see that the black strings are non-uniform. 

To see the non-uniformity of the horizon, let us investigate the change of the 
circumference radius along $y$ direction. 
The procedure to obtain a circumference radius of the horizon is summarized 
as follows: (i) seek for the radius $r=r_{+}(y)$ which satisfies 
$h_{00}+g_{00}^{(1)}=0$ 
(or equivalently $h_{11}+g_{11}^{(1)}=\infty$ as we denote below.) 
(ii) evaluate the circumference radius in the Einstein frame as 
\beqa
R_{H}:=\sqrt{r_{+}^{2}+\Psi (r_{+})g_{22}^{(1)}(r_{+})}\ .
\label{rh-bulk}
\eeqa
Note that we subtracted the effect of the AdS background in 
the above expression (See, Eq.~(\ref{holograms})). 

First, let us find $r_{+}(y)$. 
Writing $g_{00}^{(1)}=:h_{00}f_{0}(r,y)$ and $g_{11}^{(1)}=:h_{11}f_{1}(r,y)$ 
and using (\ref{correction}), 
we can verify that $f_{0}=f_{1}$ and 
they have finite values at $r=r_{H}$. 
Hence, the coordinate value of the horizon does not change 
$r_{+}(y)=r_{H}$ even at this order.  Thus, we can evaluate the 
circumference radius (\ref{rh-bulk}) using the expression   
\beqa
g_{22}^{(1)}(r_{H})=-\frac{\ell^{2}w}{2}
\left[R_{22}-\frac{1}{6}h_{22}R
+(w+2)\chi_{22}\right]\ .
\label{correction22}
\eeqa
Here, we introduced the variable $w:=\frac{1}{b^{2}}-1$ which increases 
toward the $\ominus$-brane. Notice that $R_{22}$ and $R$ are those in 
the Jordan frame. 

\begin{figure}[htbp]
\psfig{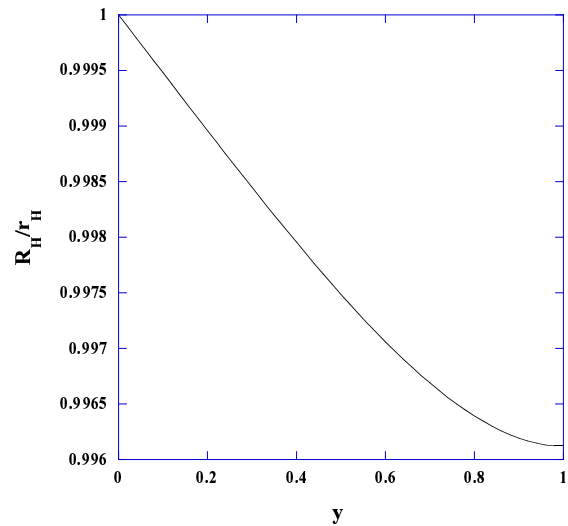}
\caption{Deformation of the horizon for $\Psi_{\infty}=0.25$, 
$\lambda_{H}=1.11\times 10^{-2}$, $\bar{a}=\sqrt{3}$ and $\ell/r_{H}=0.5$. 
\label{bulk}}
\end{figure}

As an example, we show the ratio of the horizon $R_{H}/r_{H}$ as a function of 
$y$ in Fig.~\ref{bulk}. We find that the circumference radius of the horizon 
monotonically shrinks 
toward the $\ominus$-brane even if we subtract the background effect $b(y)$. 
Although, we showed only one example, we can easily confirm that this is 
general by differentiating (\ref{correction22}) with respect to 
$w$ since the fields satisfy the conditions 
\beqa
T_{22}=\frac{\Psi Q_{m}^{2}}{2r^{2}}e^{-2a\phi}>0\ ,\ \   
T^{\mu}_{\ \mu}=0\ ,
\eeqa
at $r=r_{H}$, where $T_{22}$ and $T^{\mu}_{\ \mu}$ are those in the Jordan frame. 
In fact, by rewriting Eq. (\ref{correction22}) as 
\beqa
&&K:=-\frac{2}{\ell^{2}}g_{22}^{(1)}(r_{H}) \nonumber  \\
&&=w\left[\frac{h_{22}}{3}R+\frac{\kappa^{2}}{\Psi}T_{22}
\left\{1-\left(w+2\right)\frac{1-\Psi}{2}\right\}\right]\ ,
\label{correction22-general}
\eeqa
we obtain 
\beqa
\frac{dK}{dw}&=&\frac{h_{22}}{3}R+\frac{\kappa^{2}}{\Psi}T_{22} 
\left\{1-\left(w+1\right)(1-\Psi)\right\},  \\ 
\frac{d^{2}K}{dw^{2}}&=&-\frac{\kappa^{2}}{\Psi}(1-\Psi)T_{22}<0. 
\label{kmin}
\eeqa
 From Eq.(\ref{kmin}), we see
 $ dK /dw$ takes minimum value at the $\ominus$-brane. 
On the $\ominus$-brane, we have 
\beqa
\frac{dK}{dw}=\frac{1}{3}h_{22}R=-\frac{1}{3}h_{22}T^{\mu}_{\ \mu}=0\ .
\label{kmin2}
\eeqa
Therefore, $dK/dw$ is always positive in the bulk.
This means $ dg_{22}^{(1)} / dw <0$. As the horizon has non-trivial $y$-dependence, 
the black strings are non-uniform. 

Numerically we found that the non-uniformity of the horizon becomes larger for 
solutions with smaller gravitational mass, i.e., smaller horizon radius. 

Finally, we comment on the zeroth law of the non-uniform black string. 
Using the metric (\ref{eqn:metric}), the Hawking temperature 
$T_{H}$ is calculated as 
\beqa
T_{H}=\frac{2\pi-\kappa^{2}m_{H}'}{8\pi^{2} r_{H}}e^{-\delta_{H}}
\sqrt{\frac{1+f_{0}(r_{H},y)}{1+f_{1}(r_{H},y)}}\ ,
\label{TH}
\eeqa
Since $f_{0}=f_{1}$ at $r=r_{H}$, $T_{H}$ does not depend on $y$. 
Therefore, the zeroth law of black hole thermodynamics holds.  

\section{Conclusion }
Non-uniform black strings in the two-brane system are investigated 
using the effective action  approach.  We considered the dilaton field 
coupled to the electromagnetic field on the $\oplus$-brane. 
It is shown that the radion
acts as a non-trivial hair of the black strings. From the brane point of view,
the black string appears as the deformed GM-GHS black hole which becomes 
GM-GHS black hole in the single brane limit and reduces to the RN black hole 
in the close limit of two-branes.  In view of the catastrophe 
theory~\cite{Katz,catas,Tamaki}, our solutions are stable. 
From the bulk point of view, the black strings are proved to be non-uniform. 
Nevertheless, the zeroth law of black hole thermodynamics holds. 

\begin{figure}[htbp]
\psfig{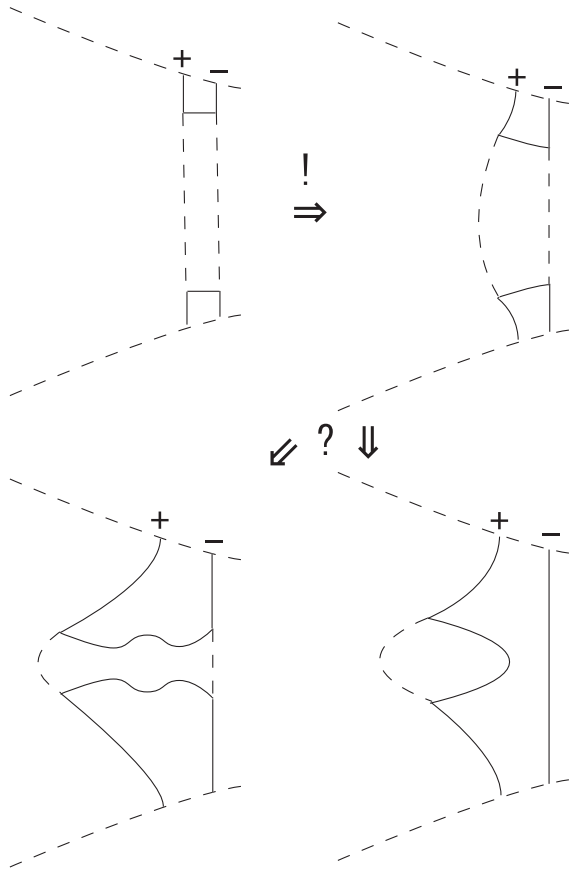}
\caption{Upper half is the image that we established in our paper. 
Lower half shows two of the possibilities when the curvature on the 
brane becomes strong. \label{image}}
\end{figure}

We established the picture shown in the upper right of 
Fig.~\ref{image}. This picture shows that the event horizon 
shrinks toward the $\ominus$-brane (even if we subtract the effect of the AdS background) 
and the distance between branes 
decreases toward the asymptotically flat region. 
As the curvature on the brane increases, the non-uniformity of the horizon becomes 
larger. 

However,  we cannot apply our present analysis if the distance between branes 
exceeds the horizon radius. This is because the Kaluza-Klein effect becomes 
significant. It is a point that Gregory-Laflamme instability commences. 
The transition to the localized black hole may occurs as in the lower right of 
Fig.~\ref{image}. The AdS/Conformal field theory correspondence argument 
suggests the classical evaporation of the resultant black hole~\cite{tanaka}.
However, the boundary condition is completely different in the two-brane system.
Therefore, the localized black hole might not be suffered from the classical 
evaporation problem. Moreover, there is also a possibility that the shape of the horizon 
becomes complicated as in the lower left of Fig.~\ref{image}. 
To get a hint, we need to proceed to the next order calculations 
corresponding to Kaluza-Klein  corrections. 
We want to investigate it in the future.

\section*{ACKNOWLEDGEMENTS}
This work was supported in part by  Grant-in-Aid for  Scientific
Research Fund of the Ministry of Education, Science and Culture of Japan 
No. 154568 (T.T.), No. 155476 (S.K.) and  No.14540258 (J.S.).  

\appendix
\section{catastrophic approach}

We proposed the stability analysis of the black holes via catastrophe theory in 
previous papers~\cite{catas,Tamaki}. We briefly summarize the discussion here. 
 The potential function of the system is described by $F(s,x)$, 
 where $s$ and $x$ are a control parameter and a state variable, 
 respectively~\cite{Poston}. The control parameter means that the system 
is controled by this parameter. The state variable is named for the reason 
that its value  is determined after the state of the system 
is determined. Thom's theorem guarantees that if a system has control parameters$\leq 4$, 
the potential function can be made to coincide with one of the 
seven elementary catastrophe's potential functions by diffeomorphism. 

The correspondence between catastrophic variables and black hole variables 
 can be identified as in Table.~\ref{class}. The equilibrium point of the system is written as 
\begin{eqnarray}
\frac{\partial F(s,x)}{\partial x}=0\ .
\label{equi}
\end{eqnarray}
In the case of the black hole system, 
this corresponds to the static solution. 
 
\begin{table}[htbp]
\begin{center}
\begin{tabular}{|l|l|}
\hline
control  & black hole mass $M$  \\
\cline{2-2}
parameters &  radion at asymptotic infinity $\Psi_{\infty}$  \\
\cline{2-2}
        & magnetic charge $Q_{m}$      \\
\hline
potential  & black hole entropy $A/4$\\
\hline
\end{tabular}
\caption{catastrophic interpretation
\label{class}}
\end{center}
\end{table}

Let us find the catastrophic criterion of the stability of static solutions. 
We write $x$ which satisfies the condition (\ref{equi}) as $x_{eq}(s)$. 
We also define $F_{eq}(s):=F(s,x_{eq}(s))$. Then we obtain 
\begin{eqnarray}
\frac{d^{2}F_{eq}(s)}{ds^{2}}&=&\left(\frac{\partial^{2} F}{\partial s^{2}}
\right)_{eq}+\left(\frac{\partial^{2} F}{\partial x\partial s}\right)_{eq}
\frac{dx_{eq}}{ds}\ .
\label{equi3}
\end{eqnarray}
Because of the condition that $F_{eq}(s)$ is an equilibrium point, we also have 
\begin{eqnarray}
\hspace{-5mm}0=\frac{d}{ds}\left(\frac{\partial F}{\partial x}\right)_{eq}=
\left(\frac{\partial^{2} F}{\partial x\partial s}\right)_{eq}
+\left(\frac{\partial^{2} F}{\partial x^{2}}
\right)_{eq}\frac{dx_{eq}}{ds}\ .
\label{equi4}
\end{eqnarray}
If we eliminate $dx_{eq}/ds$ from Eq.~(\ref{equi3}) using this 
equation, we obtain 
\begin{eqnarray}
\hspace{-5mm}\frac{d^{2}F_{eq}(s)}{ds^{2}}&=&\left(\frac{\partial^{2} F}{\partial s^{2}}
\right)_{eq}-\left(\frac{\partial^{2} F}{\partial x\partial s}\right)_{eq}^{2}
/\left(\frac{\partial^{2} F}{\partial x^{2}}\right)_{eq}.
\label{equi5}
\end{eqnarray}
Since the point where stability changes corresponds to the inflection 
point of the potential function, we have 
$\left(\frac{\partial^{2} F}{\partial x^{2}}\right)_{eq}=0$. Then, unless 
\begin{eqnarray}
\left(\frac{\partial^{2} F}{\partial x\partial s}\right)_{eq}=0\ ,
\label{equi6}
\end{eqnarray}
we have 
\begin{eqnarray}
\frac{d^{2}F_{eq}(s)}{ds^{2}}=\infty\ .
\label{equi7}
\end{eqnarray}

We apply the formula (\ref{equi5}) to black holes. 
If we fix $\Psi_{\infty}$ and $Q_m$, 
$dF_{eq}/ds$ corresponds to $d(A/4)/dM$ and equals to $1/T_{H}$ from 
the first law of black hole thermodynamics. 
Because of (\ref{equi7}), stability changes at 
\begin{eqnarray}
\frac{d(1/T_{H})}{dM}=\infty\ .
\label{equi8}
\end{eqnarray}
Therefore, what we should do is to examine if the point 
 where Eq.(\ref{equi8}) holds exists or not for various $\Psi_\infty$
 and $Q_{m}$. We have already known that the GM-GHS solutions 
are stable at least. Then, if no such a point is found, the solutions are stable 
 in the catastrophic sense. 

There may be a possibility that  (\ref{equi6}) holds. In such a case, 
the argument based on the catastrophe theory fails. 
However, the argument of the inner horizon 
supports the belief that it is not the case.




\begin{thebibliography}{99}
\bibitem{foot}
The value  $\bar{a}=\sqrt{3}$ is favored in the Kaluza-Klein theory. 
However, our explanation does not depend on the dilaton coupling qualitatively.
\end{thebibliography}
\end{document}